\def   \ni {\noindent}

\def   \ssk {\vskip  5truept}

\def   \bsk {\vskip 15truept}
\def   \newpage {\vfill\eject}
\def   \newline {\hfil\break}

\documentstyle[epsfig,rotate]{article}
\begin{document}
\hsize 5truein
\vsize 8truein
\font\abstract=cmr8
\font\keywords=cmr8
\font\caption=cmr8
\font\references=cmr8
\font\text=cmr10
\font\affiliation=cmssi10
\font\author=cmss10
\font\mc=cmss8
\font\title=cmssbx10 scaled\magstep2
\font\alcit=cmti7 scaled\magstephalf
\font\alcin=cmr6 
 \font\ita=cmti8
\font\mma=cmr8
\def\ref{\par\noindent\hangindent 15pt}
\null

\title{\ni Correlated Spectral and Temporal Variability in 
the High-Energy Emission from Blazars}
\bsk \bsk
\author{\ni C.D.~Dermer $^{1}$, H.~Li $^{2}$, J.~Chiang$^{1,3,4}$} 

\bsk
\affiliation{1) E. O. Hulburt Center for Space Research, Code 7653,
      Naval Research Laboratory, Washington, DC 20375-5352\\
\affiliation{2)  Theoretical Astrophysics, Los Alamos National
 Laboratory, Los Alamos, NM 87545\\
\affiliation{3} NRL/NRC Resident Research Asscociate\\
\affiliation{4} JILA, University of Colorado, Boulder, CO 80309-0440\\

}                                                
\bsk
\baselineskip = 12pt

\abstract{ABSTRACT \ni
Blazar flare data show energy-dependent lags and 
correlated variability between 
optical/X-ray and GeV-TeV energies,
and follow characteristic trajectories when plotted 
in the spectral-index/flux 
plane. This behavior is qualitatively explained if nonthermal 
electrons are injected over a finite time interval in the 
comoving plasma frame and cool by radiative processes.
Numerical results are presented which show the importance of the effects of 
synchrotron self-Compton cooling and plasmoid 
deceleration. The use of INTEGRAL to advance our understanding of these systems
is discussed.
}                                                    
\bsk
\baselineskip = 12pt
\keywords{\ni KEYWORDS: blazars; galaxies: jets; 
radiation  mechanism: nonthermal
}               

\bsk
\baselineskip = 12pt


\text{\ni 1. INTRODUCTION
\ssk
\ni     

Correlated X-ray and $\gamma$-ray observations  provide important
constraints on processes operating in blazars. For example,  correlated
variability is evident between the optical and 100 MeV - GeV  emission in
flat spectrum radio sources such as 3C 279 (e.g., Hartman et al.\ 1996)
and BL Lacertae (Bloom et al.\ 1997), and between the X-ray and TeV
emission in the X-ray selected BL Lac objects Mrk 421 and Mrk 501 (e.g.,
Macomb et al.\ 1995; Buckley et al.\ 1996; Catanese et al.\ 1997). The
simplest explanation is that the same population of electrons produces
both the optical/X-ray and the 100 MeV - TeV emission. Flare data for Mrk
421 (Takahashi et al.\ 1996)
 and PKS 2155-304 (Urry et al.\ 1997)
 exhibit time lags which increase with
decreasing photon energies. The Mrk 421 data  follow a well-defined
clockwise ``hysteresis'' trajectory in a spectral index/flux display, and
this behavior is also found in several other BL Lac objects (e.g., OJ
287, Idesawa et al.\ 1997; PKS 2155-304, Sembay et al.\ 1993).

These observations motivated recent studies
(Dermer 1998; Kirk et al.\ 1998) to determine whether the lags and
hysteresis diagrams can be explained through well-known processes.  Here
we briefly review these results, and  present calculations which
treat additional cooling of the electrons through the synchrotron
self-Compton (SSC) process. Plasmoid deceleration can introduce important
effects 
\newpage
\ni which resemble radiative cooling processes, and must
also be considered in detailed
 treatments  (see Chiang \& Dermer 1998;
Chiang 1998a; Dermer \& Chiang 1998). Blazar studies will be significantly
advanced using INTEGRAL, which will provide observations 
which span the X-ray/soft $\gamma$-ray regime.

\bsk
\ni 2. RADIATIVE PROCESSES
\ssk
\ni 

The simplest model for variability in blazars  (Tashiro et al.\
1995; Dermer 1998; see also Kirk et al.\ 1998, who additionally
consider electron acceleration) assumes that a nonthermal
power-law distribution of electrons is injected uniformly throughout a
relativistically moving blob over an  extended period of time, and that
the  electrons cool by synchrotron processes only. The blob is assumed 
not to accelerate or decelerate, and energy losses from
Compton scattering of photons which impinge from outside the jet 
are assumed to be small in comparison
with synchrotron losses.
The qualitative behavior of 
the energy-dependent lags and the hysteresis diagrams is reproduced. 
Moreover, by examining the energy-dependence of 
flare data at $\gamma$-ray energies, 
one can discriminate between SSC and
external Compton-scattering origins of the photons (Dermer 1998).

We (Li et al.\ 1998) have developed a code which
 includes the following radiative and cooling processes: (1) Coulomb
scattering (which is, however, generally not important);  
(2) Compton scattering and electron energy losses from arbitrary soft
isotropic photon
sources, using the full Klein-Nishina cross section; 
and (3) Cyclo-synchrotron emission, including synchrotron self-absorption and 
``synchrotron boiler''  effects on the electrons (Ghisellini et al.
1988).  The  particle injection is spatially uniform but may be
time-dependent. Particle acceleration and escape and additional soft 
photon injection is possible, though only
synchrotron photons are considered here.

The importance of the SSC process is determined by comparing the energy density 
$U_{\rm syn}$ of synchrotron photons in the comoving frame with the 
magnetic-field energy density $U_B
= B^2/(8\pi )$.  We 
can estimate the maximum value that $U_{\rm syn}$ can
 obtain by assuming that electron cooling 
has not appreciably
changed the spectrum of injected electrons at the 
end of the injection episode. If $U_{\rm syn}/U_B < 1$ 
at this moment, then the Compton cooling of the electrons
 can be neglected compared with the synchrotron cooling. 

Let ${\cal E}$ represent the energy
injected in nonthermal electrons, which we assume can be described by 
 a power law with lower and upper Lorentz factors
of the electron injection function given by $\gamma_{\rm min} $ and
$\gamma_{\rm max} $, respectively.  If we write the differential electron number spectrum
as $N(\gamma ) = N_0 \gamma^{-p}$, where $p$ is the electron injection index, then $N_0 = (2-p){\cal E}/[m_ec^2 
(\gamma_{\rm max}^{2-p}-\gamma_{\rm min}^{2-p})]$.
The value of $U_{\rm syn} \cong L_{\rm syn}/(4\pi R^2 c) \cong
[m_ec^2/(4\pi R^2 c)]\int_{\gamma_{\rm min}}^{\gamma_{\rm max}}d\gamma\; |\dot \gamma |\; N(\gamma )$, where the
electron energy loss rate through synchrotron radiation is given by $-\dot \gamma = 
[4 \sigma_{\rm T}/(3 m_e c)]U_B \gamma^2$. Thus
\begin{equation}
{U_{\rm syn}\over U_B} = {\sigma_{\rm T}\over 3\pi R^2}\;{N_0\over 3-p }\; 
(\gamma_{\rm max}^{3-p} - \gamma_{\rm min}^{3-p}) ={\sigma_{\rm T}\over 3\pi R^2}\; 
{ {\cal E}(\gamma_{\rm max}-\gamma_{\rm min})\over m_e c^2 
\ln(\gamma_{\rm max}/\gamma_{\rm min})}\;,
\end{equation}
where the last expression applies to the case $p = 2$ (see Dermer et al.\ 1997 for 
comparison with energy losses due to photons which originate from outside the jet).

In the runs shown, we let $B = 0.1$, $\gamma_{\rm min} = 10$ and
$\gamma_{\rm max} = 10^6$, and let
$p = 2$. The radius of the spherical blob is $ R = 1.5\times 10^{16}$ cm,
and the Doppler factor
${\cal D} = 10$. Using eq.\ (1), we see that $U_{\rm syn}/ U_B \simeq 0.7 {\cal E}_{46}/R_{16}^2$
for these parameters, where $R_{16} = R/(10^{16}$ cm) and ${\cal E}_{46} = {\cal E}/(10^{46}$ ergs).
We therefore choose two extreme values for ${\cal E}$
 where the effects of SSC are negligible and are dominant. In
Figs.\ 1 and 2, ${\cal E} = 10^{44}$ and
$10^{49}$ ergs, respectively. This energy in nonthermal electrons is
injected over a comoving time scale of $10^6$ s.
These figures show the evolving electron energy
distribution $E^2 n(E)$ and the  evolving
$\nu L_\nu$ photon spectrum at different times in the observer's frame. Here $n(E)= N(E)/V$ is
the differential electron energy spectrum, $V = 4\pi R^3/3$ is the blob volume,
and $E = \gamma$ m$_e$c$^2$. 

\setcounter{figure}{0}
\begin{figure}
\epsfysize=6cm
\rotate[r]{\epsffile[50 50 600 500]{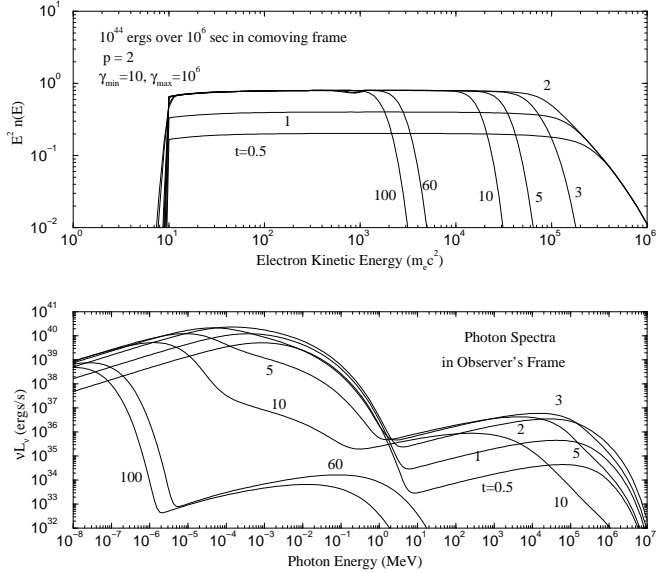}}
\caption{FIGURE 1.  Evolving electron distribution function (top panel)
and observed $\nu L_\nu$ photon spectrum (bottom panel) when $10^{44}$
ergs are injected over a time scale of $10^6$ s in the comoving frame.
The Doppler factor ${\cal D} = 10$. 
Curves are labeled by time in the comoving frame in units of the 
dynamical time scale $R/c =  5\times 10^5$ s.
}
\end{figure}

\begin{figure}
\epsfysize=6cm
\rotate[r]{\epsffile[50 50 600 500]{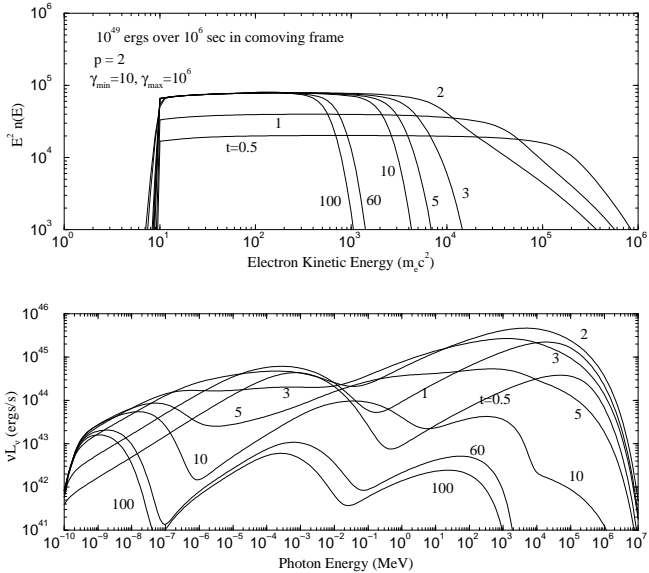}}
\caption{FIGURE 2. Same as for Fig. 1, except that $10^{49}$
ergs are injected over a time scale of $10^6$ s in the comoving frame. }
\end{figure}

The labels in Figs.\ 1 and 2 indicate time in units of the
comoving dynamical time scale $R/c = 5\times 10^5$ s.  The maximum 
intensity of the electron spectrum is therefore reached at 2 
dynamical time scales.
The evolution of the electron spectrum is obviously fastest at the highest energies.
In Fig.\ 1, where only synchrotron losses are important, the electron spectrum 
cuts off abruptly.   The synchrotron peak evolves to lower energies due to the 
electron cooling.  Even in this limit where SSC effects are not important, features from the Comptonized synchrotron emission are imprinted on the 
synchrotron spectrum, though at low levels.  By contrast, Fig.\ 2 shows the case when
SSC is very important on the electrons.  Multiple Compton scattering features are 
evident at late times in the $\nu L_\nu$ spectra. Due to the exponential escape
$\propto \exp (-ct/R)$ of the 
photons from the emitting region, high-energy ($> 10^5$ MeV) radiation is
still being received at several dynamical time scales after which electrons which 
could produce such emission have cooled.  There are no
strong  spectral softenings between hard
 X-ray and soft $\gamma$-ray energies in Figs.\ 1 and 2. We 
have also checked that compactness effects are 
 unimportant in the two calculations.

\begin{figure}
\epsfysize=6cm
\rotate[r]{\epsffile[100 -50 600 500]{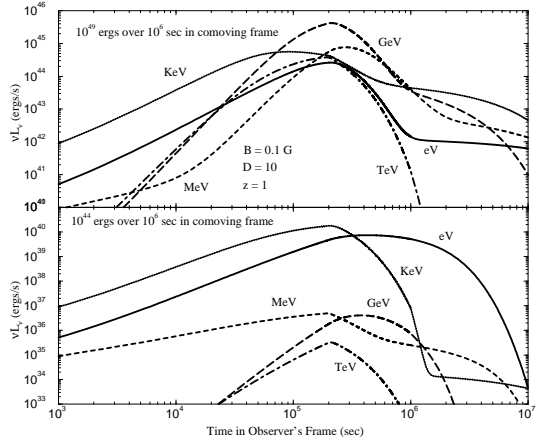}}
\caption {FIGURE 3. Light curves at eV, keV, MeV, GeV, and TeV photon energies for the 
cases where $10^{44}$ ergs (bottom) and $10^{49}$ ergs (top) are injected into the comoving frame over
a period of $10^6$ s, corresponding to the runs shown in Figs.\ 1 and 2, respectively. }
\end{figure}

\begin{figure}
\epsfysize=6cm
\rotate[r]{\epsffile[100 -50 600 400]{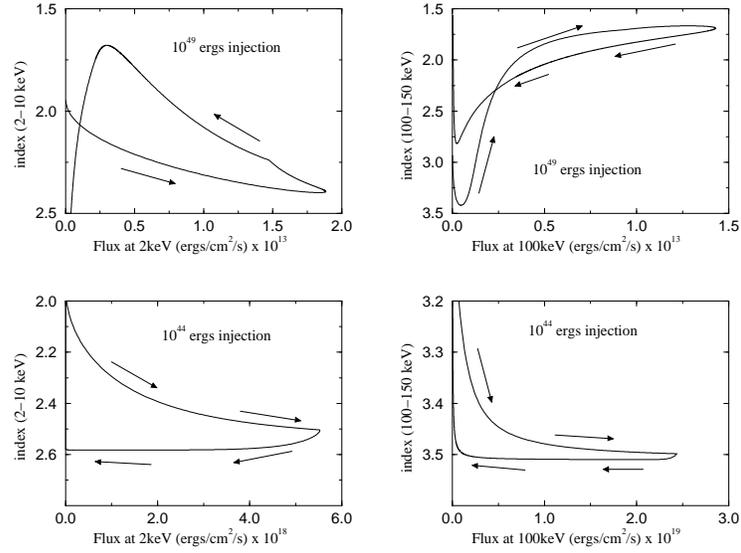}}
\caption{FIGURE 4. Model trajectories in the spectral index/flux plane at photon energies of
2 keV and 100 keV, for  the 
cases where $10^{44}$ ergs (bottom) and $10^{49}$ ergs (top) are injected into the comoving frame over
a period of $10^6$ s, corresponding to the runs shown in Figs.\ 1 and 2. }
\end{figure}

Fig.\ 3 shows light curves  calculated at different photon energies  for the 
runs of Figs.\ 1 and 2.  The light curves are calculated for a source at redshift $z = 1$.  
The comoving time is Doppler contracted and redshifted as measured by an observer; thus the
 observed dynamical
time scale $ (1+z)(R/{\cal D} c) \cong 10^5$ s.
 Consequently, these light cuves 
peak at roughly twice the dynamical time scale, or 
about 2 days. By comparing the eV and keV light curves, 
and the GeV and TeV light curves
for the $10^{44}$ ergs case, one
can see that the effects of synchrotron cooling is to
 make lower energy light curves
 lag the high energy light curves.  
In the limit where synchrotron cooling dominates, the energy dependence of the
lag is $\propto \epsilon^{-1/2}$ for the synchrotron component of the spectrum, but is $\propto \epsilon^{-1/4}$
for the SSC component (Dermer 1998). Here $\epsilon$ is the observed photon energy. When SSC losses cannot be neglected, the behavior is more 
complicated as can be seen by examining the $10^{49}$ ergs case.  Surprisingly, the keV light curves
can lag the eV light curves due to the contribution of the SSC component.

Model trajectories in the spectral index/flux plane (the so-called ``hysteresis'' diagrams)
for the flares corresponding to the runs of Figs.\ 1 and 2 are shown
in Fig.\ 4.  Clockwise rotation is found at 2 keV and 100 keV for the case where
synchrotron losses dominate the electron cooling.  When SSC losses are dominant, we
see that the hysteresis diagrams can rotate in the opposite sense at 2 keV, and exhibit even 
more complicated behavior at 100 keV.  This is 
related to the contribution of the 
SSC component which can produce a high energy component 
that leads rather than lags, as just mentioned.
Additional photon sources which impinge from outside the jet could 
produce even more complicated behavior, and remains to be studied.
\bsk
\ni
3. PLASMOID DECELERATION
\ssk
\ni
Observations of afterglows from gamma-ray bursts (GRBs) have directed attention
to the importance of extracting energy from the bulk kinetic energy of the
outflowing plasma as the mechanism for energizing the nonthermal electrons in 
the comoving frame. 
A simple review of the blast wave physics used to analyze deceleration
effects due to sweeping-up material from the surrounding medium has recently
been presented
by Dermer \& Chiang (1999).  We (Chiang 1998,1999) have 
also recently examined spectral and
temporal  effects from blob  deceleration.  
Since the characteristic electron energy in
the comoving frame is proportional to the bulk Lorentz factor of the
blob, the energy of the synchrotron peak decreases
 with time as the blob decelerates. 
This effect yields a spectral-index/flux behavior
similar in character similar to that resulting from synchrotron cooling.

In Fig.\ 5, we show a calculation using the blast-wave code developed for
GRBs by Chiang \& Dermer (1999), though with parameters appropriate
to blazars.  As shown by
Dermer \& Chiang (1999), the overall spectral properties of
blazars and GRBs are similar except for the value of their initial bulk
Lorentz factor $\Gamma_0$.  In this calculation, $\Gamma_0 = 40$, the total
energy injected per $4\pi$ sr is $10^{52}$ ergs and
the plasmoid expands into a medium with a proton
 density of 100 cm$^{-3}$. 
When it sweeps up particles, nonthermal electrons are injected with a minimum
Lorentz factor $m_p\Gamma/m_e$, and with a very soft power law slope
$p = 6$.  In this calculation, 
the magnetic equipartition parameter $\xi_B = 10^{-6}$(see
 Chiang \& Dermer 1999).

The left-hand panel of Fig.\ 5 shows the time evolution of the observed $\nu L_\nu$ spectrum.  
As time increases, the peak of the  $\nu L_\nu$ spectrum migrates from right to left due to 
the deceleration of the plasmoid. The declining magnitude of the $\nu L_\nu$ flux after it reaches its peak 
is due to plasmoid deceleration rather
than radiative cooling, which is negligible in this calculation. The right-hand panel of Fig.\ 5 shows
the trajectory followed in the spectral index/flux plane  when the plasmoid emission is observed at 2 keV.
This behavior resembles that produced by radiative cooling, 
even though radiative cooling plays essentially
no role in the variability behavior of this calculation. 
The importance of plasmoid deceleration in the 
interpretation of GRB spectral data has 
 been recently pointed out by Chiang (1999). 

Because plasmoid deceleration can produce an observed decaying flux, 
this effect must be considered whenever temporal variability observations are
used to infer properties of blazars.  It is therefore not possible to simply interpret
variability observations in terms of radiative cooling time scales.  This may weaken tests
for beaming in blazars (see Dermer 1997), or for 
inferences about the mean magnetic fields in the radiating 
plasma (e.g., Catanese et al.\ 1997; Dermer 1998).  
If variability behavior is due to plasmoid deceleration rather than to the
radiative cooling time scale, hadronic models for blazar emissions (see, e.g., Mannheim 1993)
no longer have to contend with the difficulty of demonstrating a short cooling time scale.

\begin{figure}
\epsfysize=6cm
\epsffile[-30 05 600 250]{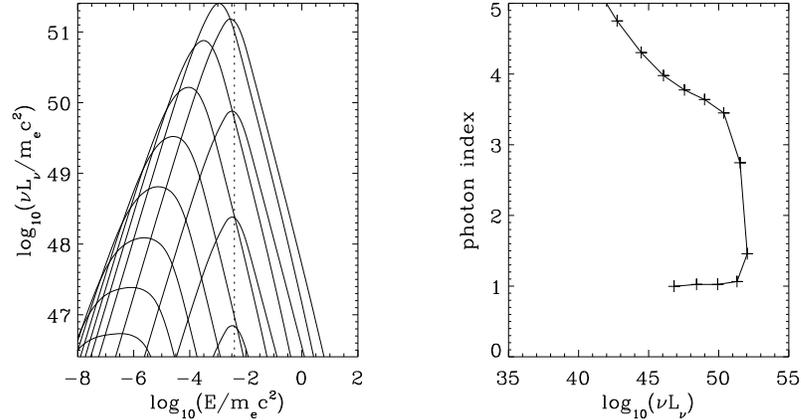}
\caption {FIGURE 5. Time-dependence of the  $\nu L_\nu$ 
spectra (left panel) and the trajectory
followed in the spectral index/flux plane at 2 keV (right panel) for a 
plasmoid which is energized and decelerates by
sweeping up material from the surrounding medium. The observer times 
when the  $\nu L_\nu$ spectra are calculated are at
$10^{j/2}$ seconds, where $j=1,2,3,\dots$. The $\nu L_\nu$ spectral peaks progress from
high to low energies with time, and the trajectory evolves from hard (photon index $\cong 1$)
 to soft. See text for other parameters. }
\end{figure}

\bsk
\ni 4. CONCLUSIONS 
\ssk
\ni 
With the launch of INTEGRAL, we can look forward to high-quality data showing
the energy-dependence of the ``hysteresis'' curves and lags over a broad
energy range from the X-ray through the soft
$\gamma$-ray regime. Correlated variability observed with the OMC on 
INTEGRAL will also contribute to studies of the variability behavior.
The calculated phase lags and trajectories are rather simply
 understood when synchrotron processes dominate the cooling,
but become significantly more complicated when the SSC process represents an 
important electron coolant, and when plasmoid deceleration is important.

Blazar science to which INTEGRAL will contribute and about which future studies
 should focus includes:
\begin{itemize}
\item The magnitude of the spectral breaks between the X-ray and soft $\gamma$-ray 
regimes, and whether this is consistent with an SSC origin of the $\gamma$ rays.
\item Phase lags as a function of photon energy, and whether higher energy emission
leads lower energy emission when the Compton component dominates.
\item Photon-energy dependent ``hysteresis'' diagrams, and under
 what conditions they can originate from radiative cooling or plasmoid deceleration, or
 exhibit counterclockwise rotation.
\item Variability time scales in the hard X-ray/soft $\gamma$-ray regime, and discriminants between
radiative cooling or plasmoid deceleration as the origin of the  variability behavior.
\item Implications of plasmoid deceleration on inferences of blazar
properties of blazars,
including the entrained magnetic field, the minimum Doppler factors, 
and the dominant radiation processes.
\end{itemize}

\bsk
\baselineskip = 12pt
{\abstract \ni ACKNOWLEDGMENTS

The work of CD is supported by the Office of Naval Research and the
{\it Compton Gamma Ray Observatory} Guest Investigator Program.
}

\bsk
\baselineskip = 12pt


{\references \ni REFERENCES
\ssk
\ref Bloom, S. D., et al.\ 1997, ApJ, 490, L145
\ref Buckley, J. H., et al.\ 1996, ApJ, 472, L9
\ref Catanese, M., et al.\ 1997, ApJ, 487, L143
\ref Chiang, J., \& Dermer, C. D. 1999, ApJ, 512, in press (astro-ph/980339)
\ref Chiang, J. 1998, ApJ, 508, 752
\ref Chiang, J. 1999, ApJ, in press (asto-ph/9810238)
\ref Dermer, C. D. 1998, ApJ, 501, L157
\ref Dermer, C. D. 1997, in Proceedings of the St. Malo INTEGRAL Workshop, 
ed. C. Winkler and T. Courvoisier, (Noordwijk: ESA), 405.
\ref Dermer, C. D., Sturner, S. J., \& Schlickeiser, R. 1997, ApJS, 109, 103
\ref Dermer, C. D., \& Chiang, J. 1999, in High Energy Processes in Accreting
Black Holes, ed. J. Poutanen \& R. Svensson, in press (astro-ph/9810222)
\ref Ghisellini, G., Guilbert, P. W., \& Svensson, R. 1988, ApJ, 334, L5
\ref Hartman, R. C., et al.\ 1996, ApJ 461, 698
\ref Idesawa, E., et al.\ 1997, PASJ, 49, 631
\ref Kirk, J. G., Rieger, F. M., \& Mastichiadis, A. 1998, A\&A, 333, 452
\ref Li, H., Dermer, C. D., \& Kusunose, M. 1998, in preparation
\ref Macomb, D. J., et al.\ 1995, ApJ, 449, L99; (e) 1996, ApJ, 459, L111
\ref Mannheim, K. 1993, A\&A, 369, 67
\ref Sembay, S., et al.\ 1993, ApJ, 404, 112 
\ref Takahashi, T., et al.\ 1996, ApJ, 470, L89
\ref Tashiro, M., et al.\ 1995, PASJ, 47, 131
\ref Urry, C. M., et al.\ 1997, ApJ, 486, 799

}                      

\end{document}